\documentclass[%
reprint,
superscriptaddress,
frontmatterverbose, 
amsmath,amssymb,aps,
]{revtex4-2}
\usepackage{xcolor}
\usepackage{graphicx}% Include figure files
\usepackage{dcolumn}% Align table columns on the decimal point
\usepackage{bm}% bold math
\usepackage{blindtext}
\usepackage{lipsum}
\usepackage{hyperref}% add hypertext capabilities
\usepackage[mathlines]{lineno}% Enable numbering of text and display math
\usepackage{amsmath}
\usepackage{multirow}
\usepackage{graphicx}
\usepackage{tabularx}
\usepackage{setspace}
\usepackage{makecell}
\usepackage{natbib}
\usepackage{todonotes}

\usepackage[%showframe,%Uncomment any one of the following lines to test 
scale=0.85, marginratio={1:1, 2:3}, ignore all,% default settings
margin=0.65in,
]{geometry}

\begin{document}

\preprint{APS/123-QED}

\title{Diffusion Driven Transient Hydrogenation in Metal Superhydrides at Extreme Conditions}% Force line breaks with \\

\author{Yishan Zhou}
   \email{have contributed equally}
   \affiliation{Center for High-Pressure Science and Technology Advance Research, Beijing, China}
    
\author{Yunhua Fu}
  \email{have contributed equally}  
  \affiliation{Center for High-Pressure Science and Technology Advance Research, Beijing, China}
  \affiliation{School of Earth and Space Sciences, Peking University, Beijing, China}
  
\author{Meng Yang}
  \email{have contributed equally}
  \affiliation{Center for High-Pressure Science and Technology Advance Research, Beijing, China}

\author{Israel Osmond}
   \affiliation{Center for Science at Extreme Conditions, Edinburgh, United Kingdom}

\author{Rajesh Jana}
  \affiliation{Center for High-Pressure Science and Technology Advance Research, Beijing, China}
  
\author{Takeshi Nakagawa}
   \affiliation{Center for High-Pressure Science and Technology Advance Research, Beijing, China}
   
\author{Owen Moulding}
    \affiliation{Institut Néel CNRS/UGA UPR2940, 25 Avenue des Martyrs, 38042 Grenoble, France} 
    
\author{Jonathan Buhot}
    \affiliation{H.H. Wills Physics Laboratory, University of Bristol, Bristol, United Kingdom}
    
\author{Sven Friedemann}
    \affiliation{H.H. Wills Physics Laboratory, University of Bristol, Bristol, United Kingdom}

\author{Dominique Laniel}
    \affiliation{Center for Science at Extreme Conditions, Edinburgh, United Kingdom}
    
\author{Thomas Meier}
  \email{thomas.meier@sharps.ac.cn}
  \affiliation{Shanghai Key Laboratory MFree, Institute for Shanghai Advanced Research in Physical Sciences, Shanghai, China}
  \affiliation{Center for High-Pressure Science and Technology Advance Research, Beijing, China}
  
\date{\today}% It is always \today, today,

\begin{abstract}
In recent years, metal hydride research has become one of the driving forces of the high-pressure community, as it is believed to hold the key to superconductivity close to ambient temperature. While numerous novel metal hydride compounds have been reported and extensively investigated for their superconducting properties, little attention has been focused on the atomic and electronic states of hydrogen, the main ingredient in these novel compounds. Here, we present combined $^{1}H$- and $^{139}La$-NMR data on lanthanum superhydrides, $LaH_{x}$, ($x = 10.2 - 11.1$), synthesized after laser heating at pressures above 160 GPa. Strikingly, we found hydrogen to be in a highly diffusive state at room temperature, with diffusion coefficients in the order of $10^{-6}~cm^2s^{-1}$. We found that this diffusive state of hydrogen results in a dynamic de-hydrogenation of the sample over the course of several weeks, approaching a composition similar to its precursor materials. Quantitative measurements demonstrate that the synthesized superhydrides continuously decompose over time. Transport measurements underline this conclusion as superconducting critical temperatures were found to decrease significantly over time as well. This observation sheds new light on formerly unanswered questions on the long-term stability of metal superhydrides.   
\end{abstract}
\maketitle
The pursuit of room-temperature superconductivity has driven significant advancements in high-pressure physics, with metal hydrides emerging as prime candidates for achieving this elusive goal\cite{Drozdov2015}. Theoretical predictions have played a crucial role in guiding experimental efforts, suggesting that hydrogen-rich compounds could exhibit superconductivity at temperatures near or even above ambient condition when subjected to extreme pressures\cite{Ashcroft2004, Zurek2009}. In particular, first-principles density functional theory (DFT) calculations predicted the formation of novel metal hydrides with complex structures and extraordinary superconducting properties\cite{Pickard2020}. These results have sparked intense research activity, leading to the synthesis of a variety of metal hydrides, including those based on sulfur\cite{Nakao2019}, yttrium\cite{Kong2019}, and lanthanum\cite{Drozdov2019}, which have demonstrated superconductivity close to ambient temperatures at high pressures.
\\
\begin{figure}[ht]
\centering    
\includegraphics[width=.7\columnwidth]{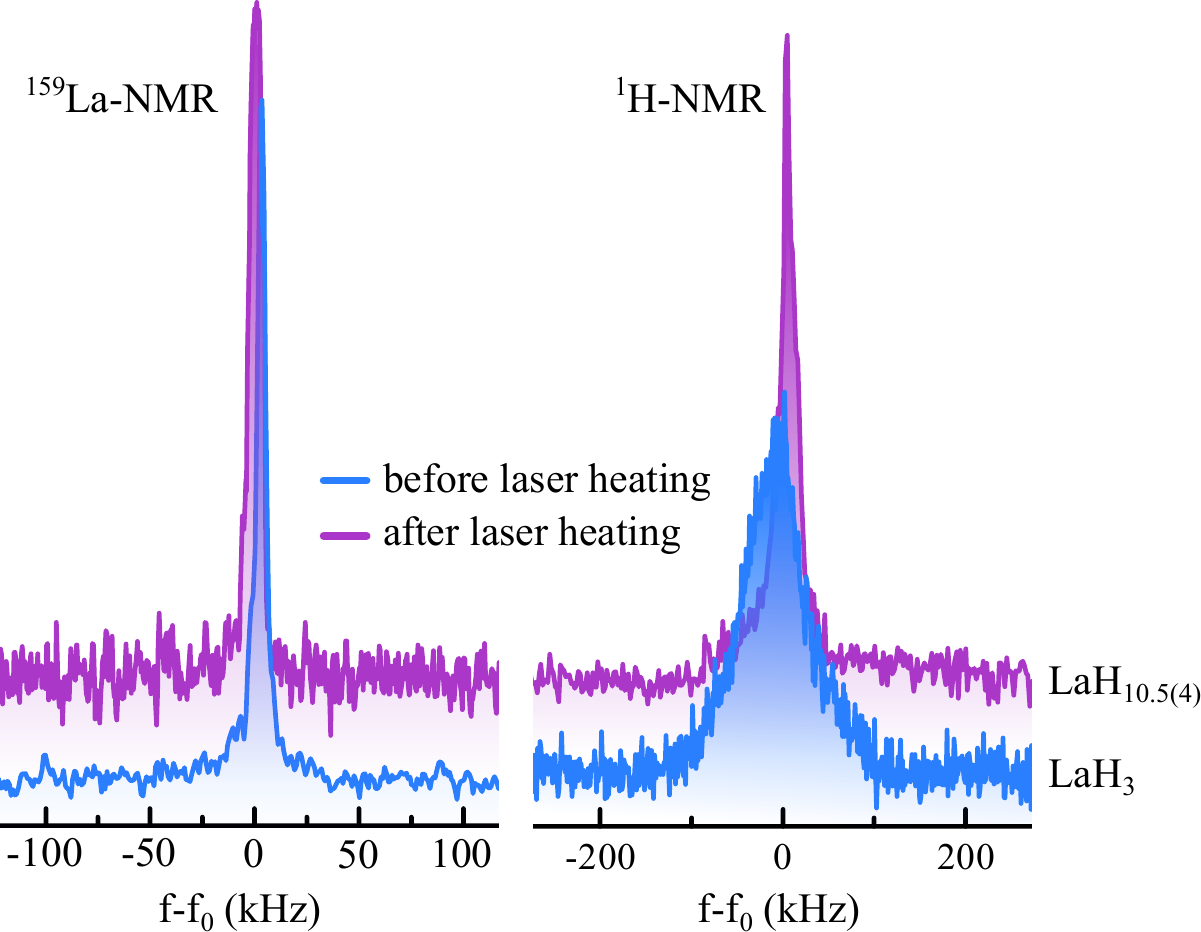}
\caption{\textbf{Lanthanum and hydrogen signals before and after laser heating above approximately 1500 K at 170 GPa.} \textbf{Left)} Sharp resonances of both $LaH_3$ and $LaH_{10.5(4)}$ indicate the absence of significant electric field gradients at the La-sites, in agreement with diffraction methods and \textit{ab-initio} DFT calculations\cite{Klavins1984X-rayTrideuteride, Chen2020a}. \textbf{Right)} While the $^1H$-NMR resonances of the hydrogen spin sub-system in $LaH_3$ are predominantly broadened by direct homo-nuclear dipole couplings, we found the corresponding resonances after laser heating to be significantly sharper, indicating enhanced hydrogen mobility\cite{Levitt2000}. Carrier frequencies $f_0$ were 55.95 MHz and 395 MHz for $^{139}La$ and $^1H$ respectively. For more information, see methods section.  
}
\label{fig:beforeafter}
\end{figure}
Among these, lanthanum superhydrides have garnered considerable attention following experimental reports of superconductivity at temperatures exceeding 250 K at pressures of around 170 GPa. Such discoveries are rooted in the idea that under sufficient pressure, hydrogen atoms in these compounds adopt an arrangement facilitating Cooper pair formation \cite{Ashcroft1968a}. Both theoretical models and the observation of the isotope effect in these systems suggest that the observed superconductivity originates from strong electron-phonon coupling, in which the hydrogen atoms provide significant  electronic density of states at the Fermi energy and high-frequency phonon modes \cite{Struzhkin2020}.
\\
However, despite these experimental and theoretical advances, significant gaps remain in our understanding of the atomic and electronic behavior of hydrogen within these dense metal hydrides. Hydrogen, due to its single electron, small interaction cross sections and high mobility, poses substantial challenges for standard high pressure experimental characterization methods. While theoretical calculations provide valuable insights into possible hydrogen configurations and the electronic structures of such materials, direct experimental probing of hydrogen's role has been limited--despite being of the utmost importance.
\\
Owing to its exceptional sensitivity to hydrogen nuclei by sensing minuscule local magnetic fields in condensed matter systems, nuclear magnetic resonance (NMR) spectroscopy has proved to be an ideal tool for investigating structural, electronic and dynamic properties of metal superhydride systems. Recent advances in the field of \textit{in-situ} high-pressure NMR led to a significant improvement of resonator sensitivities\cite{Meier2017, Meier2017b}, pressure stability\cite{Meier2018c, Meier2018b} and spectral resolutions\cite{Meier2019, Meier2021a}, allowing for routine experimentation well into the mega-bar regime\cite{Meier2020, Meier2022a, Yang2024HexagonalNMR}.
\\
Pioneering experiments on iron and copper hydride systems found a significant enhancement of the hydrogen Knight shift\cite{Knight1949} upon compression, indicating an increasing electronic density of states via the formation of a conductive sublattice of hydrogen atoms \cite{Meier2019a}. Furthermore, it could be shown that hydrogen atoms exhibit self-diffusion coefficients in the order of $10^{-8}~cm^2s^{-1}$, even at elevated pressures but room temperature\cite{Meier2020a}. Finally, developments in the field of quantitative high-pressure NMR spectroscopy\cite{Fu2024a} led to a direct and stand-alone quantification method of hydrogen atoms' concentrations in hydrides \cite{Meier2022b}.  
\\
In this context, we present an investigation addressing the aforementioned experimental and theoretical gaps by employing NMR spectroscopy to probe the atomic and electronic states of hydrogen in lanthanum superhydrides ($LaH_x$ (x = 10.2 - 11.1)) synthesized above 160 GPa. This study utilizes a combination of $^{1}H$- and $^{139}La$-NMR which reveals the highly diffusive nature of hydrogen within these compounds, which leads to a gradual de-hydrogenation of the synthesized samples over time, inferred to strongly affect the stability and superconducting properties of the synthesized superhydrides.
\\
\section{Results and Discussion}
\begin{figure}[ht]
\centering    
\includegraphics[width=.8\columnwidth]{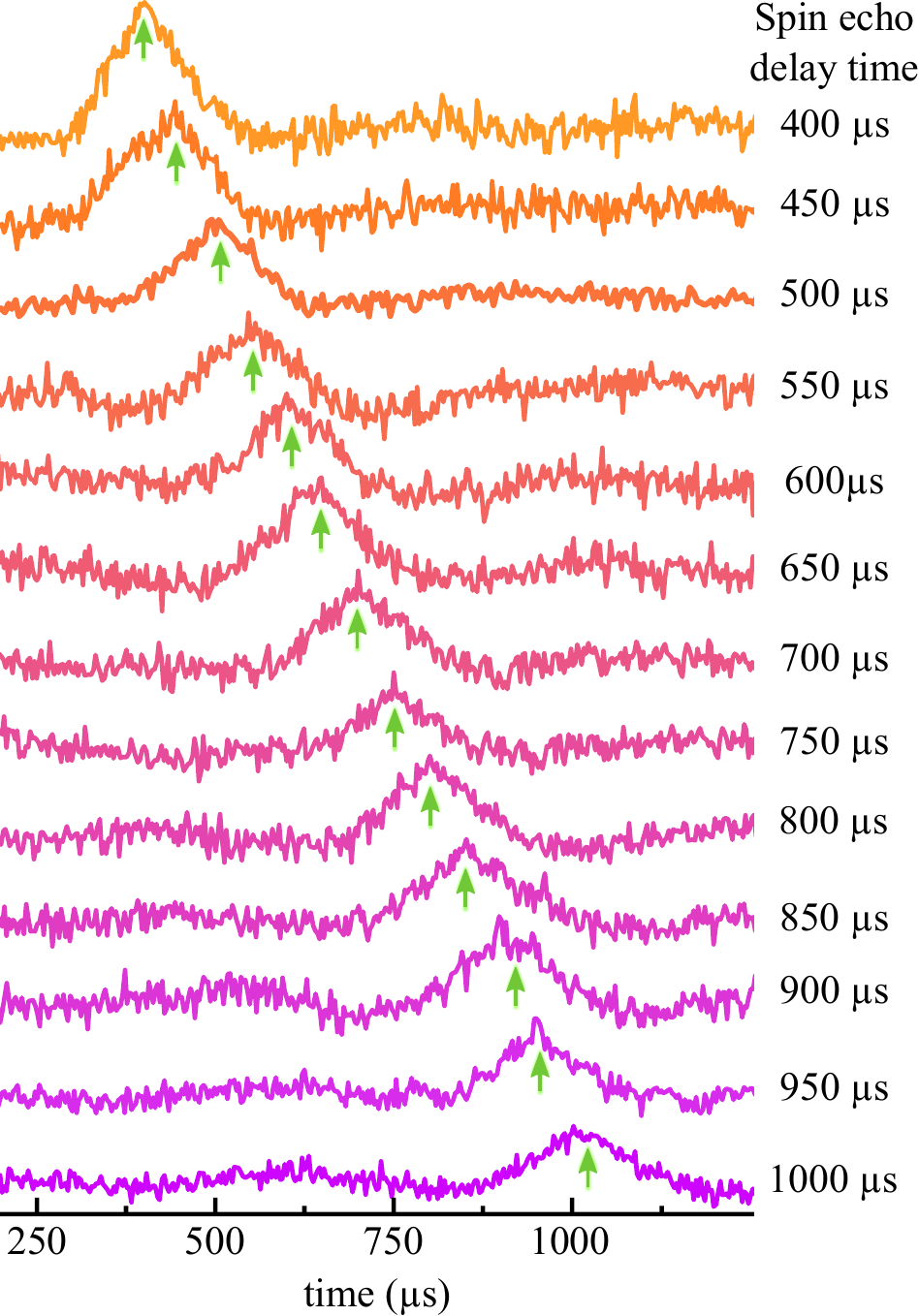}
\caption{\textbf{$^1H$-NMR time domain response after spin echo excitation in $LaH_{10.5(4)}$ at 170 GPa and room temperature.} Hydrogen spin echoes could be refocused at large delay times of up to 1 ms, indicative of greatly enhanced longitudinal spin relaxation times $T_2$.    
}
\label{fig:time domain}
\end{figure}
\begin{figure}[htb]
\centering    
\includegraphics[width=.8\columnwidth]{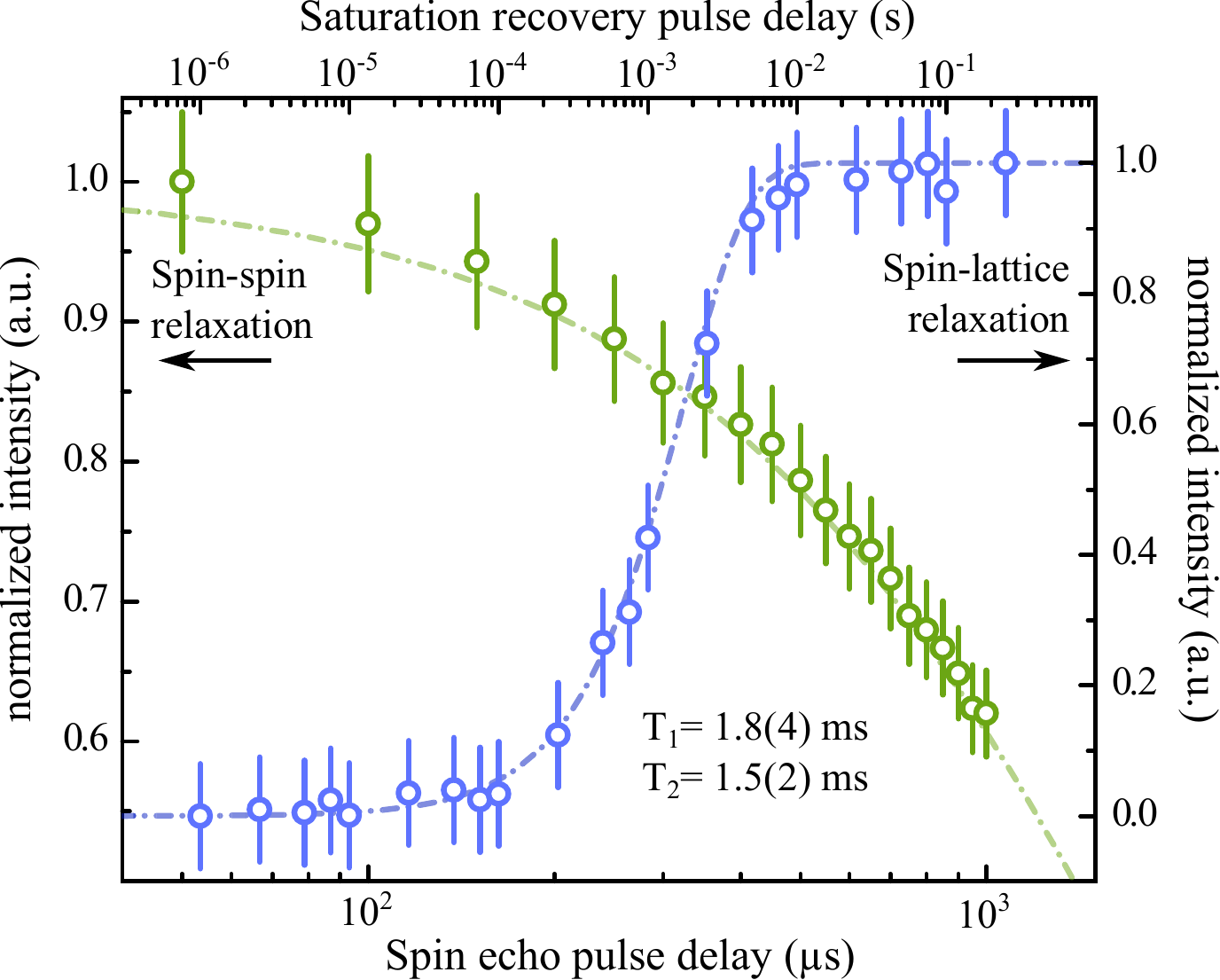}
\caption{\textbf{Comparison between spin-lattice and spin-spin relaxation times at 170 GPa at room temperature.} Both saturation recovery experiments and delayed spin echoes suggest that $T_1\approx T_2$, implying that motional correlation times of hydrogen atoms in lanthanum superhydrides approach spectral time scales ($\omega_0\tau_c\ll 1$), evidencing the highly diffusive state of hydrogen atoms in this system. Lines are respective fits for transverse and longitudinal relaxation times assuming single exponential decays.  
}
\label{fig:T12}
\end{figure}
For NMR experiments, four panoramic non-magnetic diamond anvil cells (DACs) have been prepared with diamond anvils having culets of 100 $\mu m$ in diameter. Further details are provided in the Methods section. All cells were loaded with lanthanum trihydride, $LaH_3$ and ammonia borane, $NH_3BH_3$, acting as a pressure transmitting medium as well as a hydrogen source. Samples were prepared by mixing LaH3 and NH3BH3 powders with a stoichiometric ratio of about 1:4. Synthesis of lanthanum superhydrides was achieved by laser heating the four samples above about 1500 K employing a double-sided laser heating arrangement similar to previous work\cite{Meier2019a, Meier2020a}. All NMR measurements were conducted at room temperature.
\\
Figure \ref{fig:beforeafter} shows recorded spectra of the $^{139}La$- and $^1H$-spin subsystem before and after laser heating at 170 GPa. Before laser-heating (blue spectra in Figure \ref{fig:beforeafter}), the lanthanum signal of $LaH_3$ exhibits a single sharp resonance of about 8 kHz in line-width. Given that $^{139}La$ is a $I=7/2$ quadrupolar spin system with nuclear quadrupole moments of about $20~fm^2$\cite{Harris2002NMRSpectroscopy}, making it very sensitive to local structural distortion leading to line-widths $> 100$ kHz, we conclude the precursor material to be in an almost perfect cubic environment leading to vanishing quadrupolar spin interactions\cite{Man2006} at the La lattice sites. This observation is in accordance with  X-ray diffraction data\cite{Klavins1984X-rayTrideuteride}. The hydrogen signal of $LaH_3$ was found to be broadened to about 50 kHz line-width, caused by direct homo-nuclear dipole-dipole interactions between adjacent hydrogen atoms\cite{Barnes1980ElectronicStudy.}. Microscale quantitative NMR ($\mu$Q-NMR) measurements, comparing the time-domain signal-to-noise ratios in both the $^{139}La$ and $^1H$ channels\cite{Fu2024a, Meier2022b}, found a hydrogen-to-lanthanum ratio of $H/La=3.1(2)$, in very good agreement with the anticipated stoichiometry for lanthanum trihydride. Frequency shifts for both spin sub-systems were determined to be $\sim2800~Hz/MHz$ (or ppm) and $\sim10~Hz/MHz$ for $^{139}La$ and $^1H$, respectively.
\\
\begin{figure*}[htb]
\centering    
\includegraphics[width=1\textwidth]{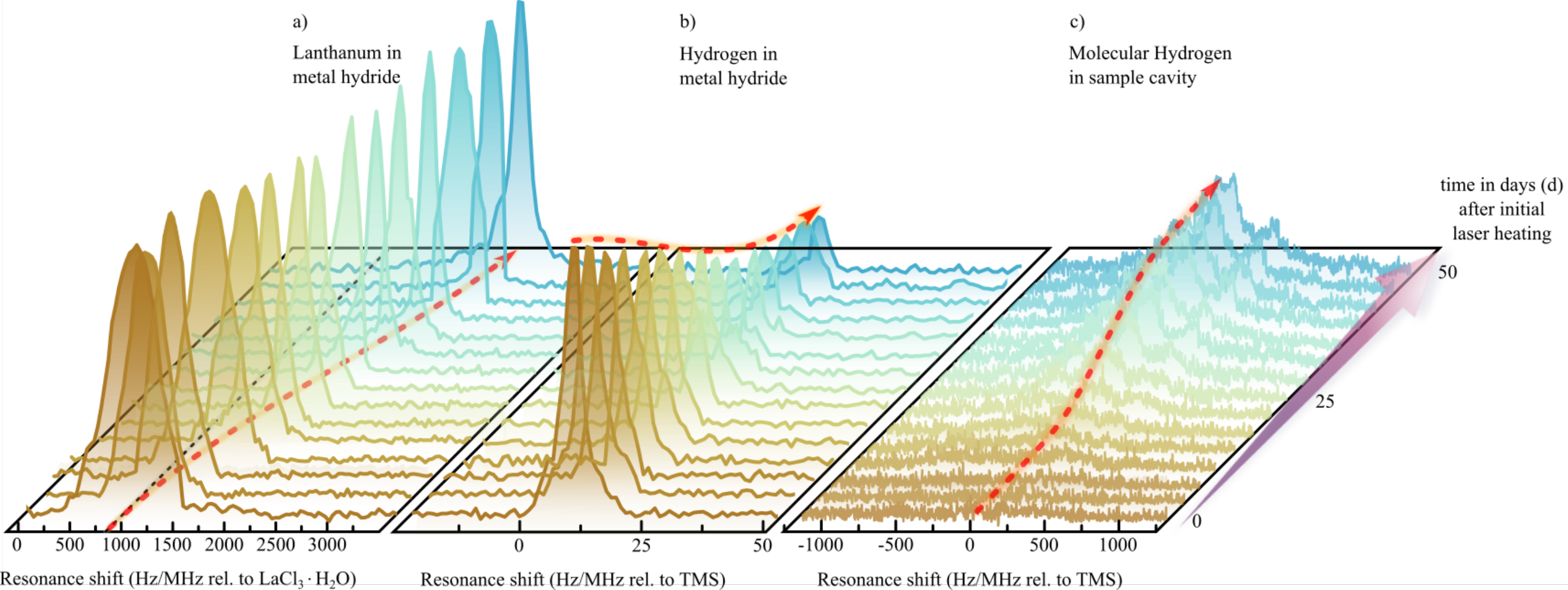}
\caption{\textbf{Long-term evolution of NMR spectra in days after initial laser heating at 170 GPa.} \textbf{a)} $^{139}La$-NMR spectra remain almost constant intensities while resonance frequencies gradually shift upfield. \textbf{b)} A progressive decrease in signal-intensity of the $^1H$-NMR spectra of the metal hydride spin sub-system was observed while a simultaneous increase of molecular hydrogen in the sample cavity (\textbf{c)}) was observed. Red lines are a guide to the eye. Carrier frequencies $f_0$ are identical to figure \ref{fig:beforeafter}
}
\label{fig:longterm}
\end{figure*}
\\ 
The purple spectra in Figure \ref{fig:beforeafter} show recorded data after laser heating. $\mu$Q-NMR measurements revealed a H/La of 10.5(4) in this particular cell. Presumably due to slightly different synthesis pressures (i.e. from 168 to 176 GPa) as well as laser-heating temperatures, hydrogen contents in all other cells varied from 10.2(3) to 11.1(4). As expected for an almost cubic arrangement of lanthanum atoms in $LaH_{10+x}$, the $^{139}La$-NMR signals were found to be very sharp, in excellent agreement with recent DFT calculations by Chen et al.\cite{Chen2020a}. 
\\
Recorded hydrogen spectra of $LaH_{10.5(4)}$ at 170 GPa exhibit strikingly sharp resonances of $\approx 1.35~kHz$ FWHM. Given that DFT-calculated average hydrogen-hydrogen distances in a compound with a face centered cubic (\textit{fcc}, space group $Fm\bar{3}m$) arrangement of lanthanum with $H/La\approx10$ are in the order of $\sim~1.0-1.1$ \AA \cite{Errea2020}, we would expect Pake-like\cite{Pake1948} resonance line-widths in the order of 90 to 120 $kHz$. Such significantly sharpened line-widths as observed here are often associated with high degrees of mobility such as in gases or liquids \cite{Slichter1978, Levitt2000}.
\\
To elucidate the possibility of enhanced mobility of hydrogen atoms in this lanthanum superhydride, we recorded several spin echo trains with increasing de-phasing times\cite{Hahn1950}. Figure \ref{fig:time domain} shows recorded hydrogen spin echoes at 170 GPa. As it can be seen, echoes are observable at delay times of up to 1 ms and beyond. In general, the transverse relaxation ($T_2$) in condensed matter solid-state systems tends to become shorter under pressure, e.g., on the order of $20-50~\mu s$, whereas our experiments indicate a spin-spin relaxation of $T_2=1.5(2)~ms$. Additionally, we conducted saturation recovery experiments in order to determine the longitudinal spin-lattice relaxation time ($T_1$) of the hydrogen spin sub-system, see Figure \ref{fig:T12}. Strikingly we found that both $T_1$ and $T_2$ are almost identical, evidencing that correlation times of atomic motion of the hydrogen atoms to be much faster than spectral time scales $1/\omega_0=1/(2\pi f)\approx0.4~ns$. In this so-called extreme narrowing limit ($\omega_0\tau_c\ll1$) both relaxation times converge, leading to an effective cancellation of broadening spin interactions like homo-nuclear dipole-dipole interaction\cite{Bloembergen1948, Hanabusa1966, Abragam1961}.
\\
Indeed, in the limit of extreme narrowing, correlation times of motion -and thus atomic diffusion coefficients- become directly proportional to the transverse spin-lattice relaxation time via\cite{Abragam1961, Meier2020a}: 
\begin{equation}
    D=\frac{3\pi}{10}\cdot\frac{\gamma_n^4 h^2 N_0 T_1}{a}
\end{equation}
with $\gamma_n$ the gyromagnetic ratio, $N_0$ the atomic number density, and $a$ the shortest contact between between two hydrogen atoms. Thus, for a number density of about $N_0\approx3\cdot10^{23}~cm^{-3}$ and shortest H-H distance being on the order of $1$ \AA, the diffusion coefficients of hydrogen atoms in $LaH_{10+x}$ are found to be about $D\sim10^{-6}~cm^2s^{-1}$. Recent \textit{ab-initio} calculations by Causs\'e $et$ $al.$\cite{Causse2023} predict similar diffusion coefficients in lanthanum superhydrides, albeit at higher temperatures. We speculate that the strong quantum fluctuations relevant for the stability of LaH$_{10}$ \cite{Errea2020} give rise to the large diffusion coefficient for hydrogen at room temperature.
\\
\begin{figure}[ht]
\centering    
\includegraphics[width=.9\columnwidth]{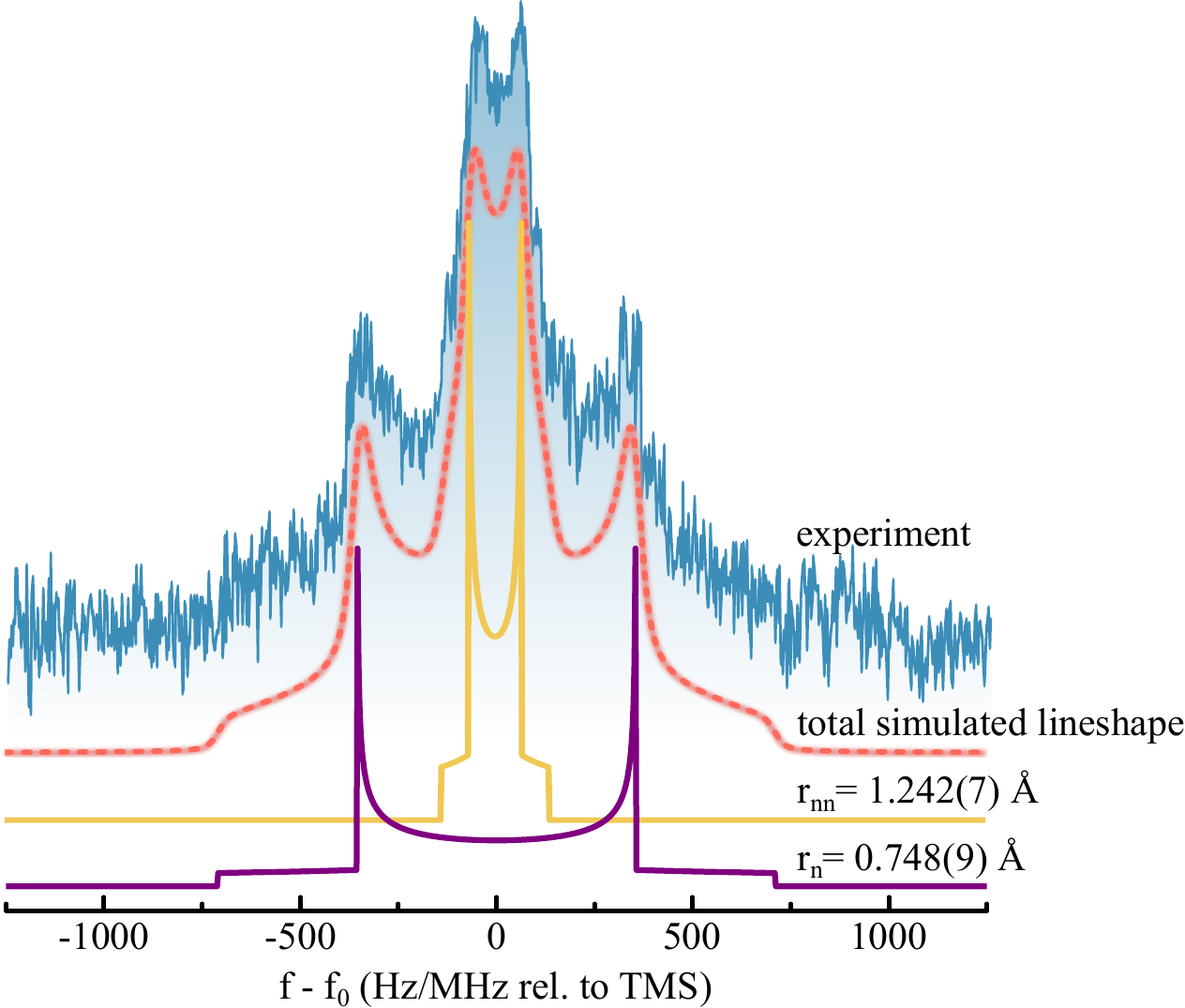}
\caption{\textbf{$^1H$-NMR spectrum of molecular hydrogen detected 54 days after initial laser heating at 170 GPa.} The observed line-shape is found to be a superposition of two overlapping dipole-dipole spectra associated with first($r_n$)- and second-nearest ($r_{nn}$) hydrogen distances in hydrogen phase III.
}
\label{fig:phaseIII}
\end{figure}
Experiments at ambient conditions suggested that highly diffusive hydrogen in metal hydrides might play a significant role in their observed decomposition over longer time periods\cite{Potzel1984, Zogal1975}. Indeed, diffusion-driven hydrogen desorption is a commonly known effect in hydrogen storage materials and low-H content hydrides at ambient conditions\cite{Sakintuna2007MetalReview}. 
\\
In order to examine the effect of hydrogen desorption, we measured both $^{139}La$- and $^1H$-NMR signals of all prepared four diamond anvil cells over a time period of 50 to 70 days without otherwise changing experimental conditions. Pressures were checked on a daily basis, without any significant changes -within 15 GPa- over the investigated time periods. Figure \ref{fig:longterm} shows the recorded lanthanum and hydrogen signals in the sample cavity. 
\\
Directly after the synthesis of the high hydrogen content compounds through laser-heating, the dominant $^{139}La$ signal was found at Knight shifts of about 900 Hz/MHz relative to an aqueous solution of $LaCl_3$ , Figure \ref{fig:longterm}a. We observed an increase in the FWHM line-width to about 38 kHz, indicating slightly distorted local lanthanum environments and an onset of first order quadrupole interaction. However, no quadrupolar satellite signals of higher-order quantum transitions could be observed, indicating the lanthanum sites remain in isotropically cubic local environments. Remarkably, over time, the lanthanum NMR signals were found to gradually shift down-field (towards higher shift values), whereas overall signal intensities -and thus the number of lanthanum atoms in the sample cavity- remained unchanged. About 50 days after the initial sample synthesis, we found the $^{139}La$-NMR signal to be shifted to 2950 Hz/MHz, which is only about 150 Hz/MHz higher than the observed Knight shift of the LaH$_3$ precursor material.
\\
\begin{figure*}[ht] 
\centering    
\includegraphics[width=.85\textwidth]{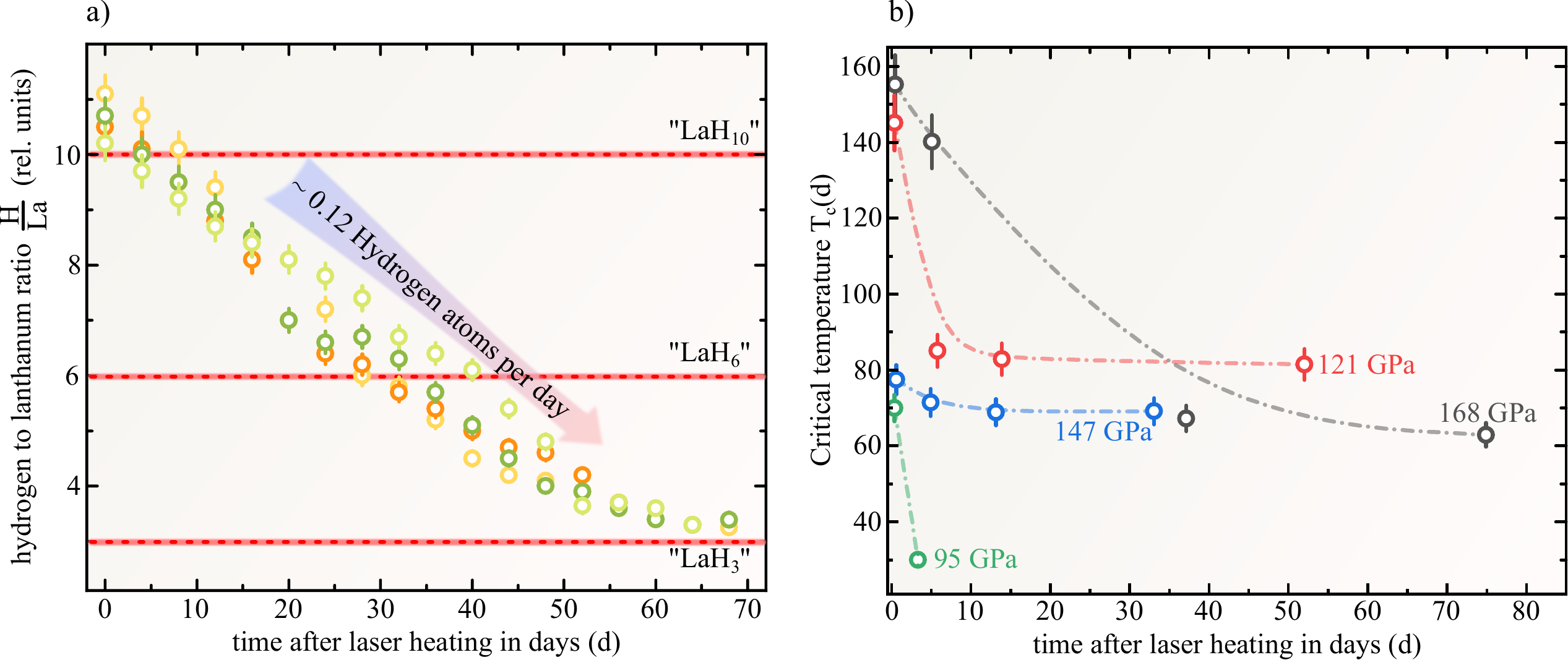}
\caption{\textbf{Long-term evolution of hydrogen contents in metal hydride compound and superconducting critical temperatures.}
\textbf{a)} Evolution of the hydrogen-to-lanthanum ratio $\frac{H}{La}$ over the course of 68 days summarizing data of all four employed DACs at room temperature. $\frac{H}{La}$ was found to continuously decrease with time, asymptotically approaching the hydrogenation of precursor $LaH_3$, with a desorption-rate of about 0.12 hydrogen atoms per day. \textbf{b)} Evolution of the normalised superconducting critical temperature $T_c$ after initial laser heating. Four DACs were laser heated at different pressures ranging from 95 to 168 GPa. As can be seen, all synthesised compounds express a significant decrease of $T_c$ over time.
} 
\label{fig:overtime}
\end{figure*}
At the same time, we observe a gradual loss of signal intensity in the $^1H$-NMR spectrum associated with the synthesized $LaH_{10+x}$ samples, as shown in Figure \ref{fig:longterm} b). No significant resonance shift could be observed. Recorded spectra suggest an increase of FWHM line-widths as time progresses. Furthermore, the formation of a broad signal of about $\sim10^3$ Hz/MHz was observed, as shown in Figure \ref{fig:phaseIII}. This gradually appearing signal has striking similarities with recently observed $^1H$-NMR signals of molecular hydrogen phase III\cite{Yang2024HexagonalNMR} after the collapse of the spin isomer distinction\cite{Meier2020}. If this spectrum is indeed interpreted in terms of a superposition of two Pake doublets broadened by dipolar interactions\cite{Pake1948} between first($r_n$)- and second-nearest ($r_{nn}$) hydrogen distances we find $r_n=0.75(1)$ \AA~and $r_{nn}=1.24(2)$ \AA, in excellent agreement with \textit{ab-initio} calculations by Labet $et$ $al.$ of hydrogen phase III\cite{Labet2012}. Thus, we conclude the observed broad signal to be associated with molecular hydrogen.
\\
The observation of a decrease in intensity of the $^1H$-NMR signal associated with the synthesized lanthanum hydride, and the concurrent formation of molecular hydrogen in the sample cavity, indicates that hydrogen continuously desorbs out of the $LaH_{10+x}$ across the pressure range studied.
In order to illustrate this hypothesis, we conducted $\mu$Q-NMR experiments for each data collection. Figure \ref{fig:overtime}a shows the resulting H/La ratios over the entire experimental observation period.
\\ 
As can be seen, directly after laser heating, H/La ratios were found to be between 10.2(3) and 11.1(3) for synthesis pressures ranging from 168 to 176 GPa, close to previously reported hydrogenations in lanthanum superhydrides\cite{Geballe2018, Drozdov2019, Laniel2022}. A continuous decrease in overall H/La atomic ratios in the metal hydride systems was found, with an average hydrogen loss of 0.12(5) atoms per day.
\\
The horizontal red lines in Figure \ref{fig:overtime}a indicate H/La ratios associated with lanthanum hydride stoichiometries of $LaH_x$(x = 10, 6, 3). It was found that after about 30 days after initial laser heating, the overall hydrogen to lanthanum ratio decreased by almost 40\%, indicating a gradual continuous transition from stoichiometries associated with $LaH_{10}$ to $LaH_6$.
\\
At observation periods longer than 50 days, data suggests a somewhat flattened out convergence to atomic ratios reminiscent of the lanthanum trihydride precursor. As this convergence remains poorly resolved in the given data (Figure \ref{fig:overtime}a), observations at longer time intervals would be more indicative of whether a complete decomposition to the precursor materials is taking place. Unfortunately, measurements at longer times were not possible due to diamond failure caused by hydrogen diffusion and consequent embrittlement. However, hydrogen absorption and desorption studies on magnesium dihydride suggest that dynamic dehydrogenation over time converges to slightly higher H-contents than that of the precursor.\cite{Huot1999StructuralHydride}.
\\
For a clathrate hydride system in which high hydrogen content is expected to be beneficial for superconductivity, a decrease in the hydrogen content should result in a significant reduction of superconducting critical temperature as a function of time. Figure \ref{fig:overtime} b), shows four different experimental runs -determining $T_c$ via transport measurements- conducted in four DACs at different synthesis pressures (see Materials section), obtained over a time of up to 80 days. Laser heating produces $Tc$ values up to 160K observed immediately after laser heating. While the structure and stoichiometry of such initially synthesized hydride phases are unknown, $T_c$ was found to drop sharply in the days following laser heating, before converging to significantly lower values between 70 and 85 K, well below that expected for $Fm\bar{3}m$ LaH$_{10}$ of $>$ 240K within this pressure range\cite{Drozdov2019}. 
\\
Given the here presented data, the following picture emerges. Directly after laser heating, lanthanum atoms occupy lattice sites resulting in vanishing electric-field gradients, in accordance with DFT structural search algorithms and diffraction experiments\cite{Geballe2018,Laniel2022,Drozdov2019}. Our NMR data clearly shows the hydrogen atoms to be in a state of high atomic mobility, with diffusion coefficients in the order of $10^{-6}~cm^{2}s^{-1}$, several orders of magnitude more mobile than in known hydrides at ambient conditions\cite{Majer1995, Wipf2000}. This diffusive state of hydrogen leads, over time, to a dynamic desorption of hydrogen out of the synthesized hydrides, accompanied by a continuous release of molecular hydrogen phase III in the sample cavity. Using quantitative NMR methods, it could be shown that over the course of about 70 days, initially synthesized LaH$_{10-11}$ samples approached a state reminiscent of almost full decomposition back to the precursor materials. Likewise, for the transport cell with the highest synthesis pressure (168 GPa), we observe a timescale for the decrease and stabilization of $Tc$ comparable with the desorption timescale observed in NMR experiments at a similar synthesis pressure. Given differing sample geometries between NMR and transport experiments, a quantitative comparison of hydrogen diffusivities is however difficult and likely also depends on the relative amounts of excess molecular H$_2$ produced from the decomposition of $NH3BH3$, the amount of synthesized LaH$_x$ and the hydrogen porosity of the gasket material itself. 
\\
At this point it should be noted that NMR measurements in diamond anvil cells often require different and somewhat simpler sample geometries than usual experimental setups used for transport experiments. Whereas transport measurements often require a more stringent sample geometry in order to fully bridge electrical contacts, NMR measurements are somewhat more flexible. Loading the sample cavity using finely ground powder of about $>1~\mu m$ grain sizes is preferable to ensure proper powder averaging of orientation dependent spin interactions like homo-nuclear dipole-dipole couplings or quadrupolar interactions\cite{Zheltikov2006}. 
\\
Such fine dispersed powders are much more effectively hydrogenated upon the release of H$_2$ from NH$_3$BH$_3$ to form LaH$_{10}$, whilst the observed initial T$_c$ in transport evidences only partial hydrogenation. The observation of a full superconducting transition is sensitive only to a continuous superconducting path between the electrical contacts, of which the highest T$_c$ superconducting transition will dominate the observed resistance behavior. From the above transport data, we conclude that high-T$_c$ lanthanum hydride phases with T$_c$ values above 140 K are not stable for our presented measurements.  
\\
Furthermore, it needs to be mentioned that NMR probes the entire sample cavity in order to pick up faint nuclear induction signals after radio-frequency excitation. Thus, provided that this method relies greatly on the number of active NMR nuclei (nuclear spin $I>0$), small grains of different stoichiometries might be below our detection limits. This is likely also the reason why observed NMR spectra, e.g., Figures \ref{fig:beforeafter} and \ref{fig:longterm}, only show single signals, despite commonly accepted structural heterogeneity in DAC sample cavities after laser heating\cite{Laniel2022}. During sample synthesis, this heterogeneity has been tried to mitigate by de-focusing of laser spots, long-term exposure, as well as careful rastering over the entire sample cavity. Nevertheless, we cannot rule out the existence of different hydride systems in the sample cavities after laser heating, but given that all $\mu$Q-NMR analyses indicated atomic ratios of H/La $>10$ we assume that $LaH_{10+x}$ is the most dominant hydride phase synthesized.   
\\
The here observed phenomenon has significant implications for studies of hydride materials at high pressures. The desorption from higher hydrides like LaH$_{10}$ is likely driven by the lack of a sufficient hydrogen reservoir in our experiments and most likely effects other experimental approaches. The absence of excess hydrogen is apparent in our experiments from the absence of an NMR signal of molecular hydrogen immediately after the synthesis of the NMR samples and the absence of a H$_{2}$ vibron in Raman measurements of our electrical transport samples. The desorption of hydrogen from the samples suggests that the chemical equilibrium with fully hydrogenated LaH$_{10}$ may only be stable in the presence of excess hydrogen. This aspect should be taken into account to design future studies of hydrides at high pressures. It will also be interesting to check if the desorption can be halted at low temperatures as suggested by recent transport studies on La$_{4}$H$_{23}$ \cite{Cross2024}. We note that the desorption and the chemical instability might explain the large sample dependencies observed for higher lanthanum hydrides \cite{Somayazulu2018, Drozdov2019} and might similarly apply to yttrium hydrides \cite{Kong2021}. By contrast, sulphur hydride (H$_{3}$S) samples show a stable T$_{c}$ over years and lack large sample dependencies \cite{Drozdov2015,Osmond2022Clean-limit/mrow}. Understanding the differences in hydrogen diffusion and desorption in hydrides might pave a way to stabilise hydrides at lower and ambient pressure.
\\
In summary, this work represents a novel viewpoint on metal hydrides under extreme conditions. It could be shown that the highly diffusive state of hydrogen atoms leads to transient hydrogen contents with highest values of H/La as well as highest values of $T_c$ occurring in a narrow time frame directly after sample synthesis followed by steep declines in both quantities. This observation sheds new light on ongoing experimental controversies and opens up new questions on the theoretical treatment of potentially superconducting metal hydrides synthesised in diamond anvil cells. 
%%%%%%%%%%%%%%%%%%%%%%%%%%%%%%%%%%%%%%%%%%%%%%%
\section*{Methods}
\textit{NMR-DAC preparation}
\\
DACs for high-pressure NMR experiments were prepared following a procedure described in previous references\cite{Meier2017,Meier2019a}. First, rhenium gaskets were indented to the desired thickness, usually $\lesssim20 \mu$m. Sample cavities were drilled using specialized laser drilling equipment. After gasket preparation, the diamond anvils were covered with a layer of 1 $\mu$m of copper or gold using chemical vapour deposition. To ensure electrical insulation of the conductive layers from the rhenium gasket, the latter were coated by a thin layer ($\approx 500$ nm) of Al$_2$O$_3$ using physical vapour deposition. The Lenz lens resonators were shaped from the conductive layer on the diamonds by using focused ion beam milling. 
\\
Before the final cell assembly, radio frequency resonators were prepared accordingly to their desired operation frequency. Pairs of high inductance solenoid coils ($\approx 100$ nH) for $^{139}La$ and $^{1}$H-NMR frequencies at high fields were used as driving coil arrangements for the Lenz lens resonators' structure and were placed around each diamond anvil. After sample loading and initial pressurisation, the driving coils were connected to form a Helmholtz coil-like arrangement.
\\
Pressure was determined using the shift of the first derivative of the first order Raman signal of the diamond edge, collected in the center of the anvils' culet. All DACs were fixed and connected to home built NMR probes equipped with customized cylindrical trimmer capacitors (dynamic range of $\approx20$ pF) for frequency tuning to the desired resonance frequencies and impedance matching to the spectrometer electronics ($50~\Omega$). 
\\
Proton and lanthanum shift referencing were conducted using the $^{63}$Cu resonances of the Lenz lenses themselves as internal references taking into account the additional shielding of $B_0$ inherent to every DAC. These resonances were cross referenced with standard metallic copper samples at ambient conditions without a DAC. The resulting shift between both $^{63}$Cu-NMR signals are then used as a primer for the NMR signals of the samples under investigation. Additionally, two DACs were prepared in the same way as described above and loaded with an aqueous solution of $LaCl_3$ for lanthanum referencing and with tetramethylsilane for hydrogen shift referencing. Both methods yielded similar results, with shifts varying by less than 5 ppm.
\\
After pressurisation to pressures above 160 GPa, double-sided laser heating to temperatures above 1200 K was achieved. In order to ensure a roughly homogeneous sample environment, laser heating was performed with defocused laser spots and the sample rastered several times. Resulting NMR signals did not show more than single signals above the detection limit.  
\\
\textit{NMR experiments}
\\
All NMR experiments were conducted on a modified 395 MHz proton frequency magnet, equivalent to a magnetic field of 9.28 T, using fully homemade NMR probes. $^1$H-NMR spectra of lanthanum hydrides were recorded using a single pulses with Gaussian shape modulation. Spin echo spectra of molecular hydrogen phase 3 were recorded using a Hahn-echo sequence with Gaussian pulse modulation. All lanthanum spectra were recorded using solid echo pulse sequences with Gaussian modulation. Atomic ratios of H/La were determined using the method of time-domain quantification similar to Fu $et~al.$\cite{Fu2024a}
\\
\textit{Transport Cell Preparation}
\\
Electrical transport measurements were carried out using diamond anvil cells equipped with diamonds with 50 µm culet. Gaskets were made from T301 stainless steel, preindented to a thickness below 40 µm. The centre of the steel was then drilled out and replaced with a mixture of cubic-boron nitride and epoxy mixture to insulate electrodes from the gasket. Electrodes were formed by thin-film deposition of six tungsten and gold bilayers, followed by the deposition of lanthanum samples (20 x 20 x 0.2 $\mu$m) directly onto the electrodes at the centre of the culet. 
\\
Cells were  closed with $NH_3BH_3$, employed as both the pressure-transmitting medium and the hydrogen source. After pressurisation to a given target pressure, these cells were laser heated to 1200 K using a 1070 nm YAG laser with multiple 300 ms pulses. After heating, resistance measurements as a function of temperature were immediately performed, and then periodically over subsequent days to weeks using an AC-resistance bridge (SIM921, Stanford Research Systems) with a 100 $\mu$A excitation current.
\\
\section*{Data availability}
The data supporting the findings of this study are publicly available from the corresponding author upon request.
\\
\section*{Code availability}
Python code used for analysing NMR data is available from the corresponding author upon request.
\\

%\bibliographystyle{unsrtnat}
%\bibliography{references.bib}

\section*{Acknowledgements}

We would like to acknowledge fruitful discussions with Ho-kwang Mao and Yang Ding. This work was supported by the National Science Foundation of China (42150101) and the National Key Research and Development Program of China (2022YFA1402301). T. Meier acknowledges financial support from the Center for High Pressure Science and Technology Advanced Research as well as the Shanghai Key Laboratory \textit{MFree} and the Institute for Shanghai Advanced Research in Physical Sciences, Pudong, Shanghai. D.L. thanks the UKRI Future Leaders Fellowship (MR/V025724/1) for financial support. For the purpose of open access, the authors have applied a Creative Commons Attribution (CC BY) licence to any Author Accepted Manuscript version arising from this submission.  
\\
\section*{Author Contributions Statement}
Idea and Conceptualisation: T.M., Y.Z., M.Y.; Methodology:  T.M. and Y.Z.; Investigation: M.Y., Y.Z, R.J., T.N., I.O.,D.L., S.F.,O.M., J.B., Y.F and T.M.; Visualization: Y.Z., and T.M.; Formal analysis: Y.Z., T.M.; Validation: M.Y., Y.Z., R.J., I.O., T.M.,I.O.,D.L., S.F.,O.M., J.B.; Funding acquisition: T.M., S.F., D.L.; Resources: T.M., S.F., D.L.; Project administration: T.M.; Writing—original draft: Y.Z. and T.M.; Writing—review and editing: M.Y., Y.Z., R.J., T.N., Y.F, S.F., D.L., O.M., J.B., I.O. and T.M.

\section*{Competing Interests Statement}

We declare no competing interest.

\end{document}